# Quantum Magnetic Properties in Perovskite with Anderson Localized Artificial Spin-1/2

*Jagath Gunasekera, Ashutosh Dahal, Yiyao Chen, Jose A. Rodriguez-Rivera,*
*Leland W. Harriger, Stefan Thomas, Thomas W. Heitmann, Vitalii Dugaev, Arthur Ernst,*
*and Deepak K. Singh\**

Quantum magnetic properties in a geometrically frustrated lattice of spin-1/2 magnet, such as quantum spin liquid or solid and the associated spin fractionalization, are considered key in developing a new phase of matter. The feasibility of observing the quantum magnetic properties, usually found in geometrically frustrated lattice of spin-1/2 magnet, in a perovskite material with controlled disorder is demonstrated. It is found that the controlled chemical disorder, due to the chemical substitution of Ru ions by Co-ions, in a simple perovskite $CaRuO_3$ creates a random prototype configuration of artificial spin-1/2 that forms dimer pairs between the nearest and further away ions. The localization of the Co impurity in the Ru matrix is analyzed using the Anderson localization formulation. The dimers of artificial spin-1/2, due to the localization of Co impurities, exhibit singlet-to-triplet excitation at low temperature without any ordered spin correlation. The localized gapped excitation evolves into a gapless quasi-continuum as dimer pairs break and create freely fluctuating fractionalized spins at high temperature. Together, these properties hint at a new quantum magnetic state with strong resemblance to the resonance valence bond system.

## 1. Introduction

The quantum spin fluctuation is key to the understanding of the microscopic mechanism behind the novel quantum critical phenomenon or the quantum spin liquid state.[1–3] The quantum spin liquid state arises from the phase coherent quantum fluctuation of entangled spin-1/2 with net spin of

0 and 1 $\mu_B$.[4] Depending on the nature of the entanglement, whether it is an ordered partitioning of nearest neighbor spins or the superposition of an infinite partitioning of nearest as well as further away spins that are simultaneously pointing along different directions, the collective ensemble of dimers can be characterized as the valence bond solid (VBS) or the valence bond liquid (VBL), respectively.[5–7] Since the moment is in continuous quantum fluctuation mode, the singlet state is either excited to the triplet state or the valence bond breaks and a continuum excitation due to the fluctuation of individual spin yields a liquid-like gapless state.[5,8] An inherent frustration in the disorder free triangular lattice plays an important role in the realization of these novel effects. However, creating a disorder free magnetic lattice is a difficult task to achieve.[9,10]

Taking an altogether different route, we have explored the quantum mechanical properties of spin-1/2 in a disorder induced tailored frustrated system (illustrated in **Figure 1**a). A quenched disorder prohibits the propagation of long-range static or dynamic order, hence automatically creates frustration. The localized impurity, due to the chemical substitution of Ru by Co-ions in a well-defined lattice of $CaRuO_3$ perovskite,[11,12] is analyzed using the Anderson localization scheme (see Section 2.2). First principle

Dr. J. Gunasekera, A. Dahal, Dr. Y. Chen, Prof. D. K. Singh
Department of Physics and Astronomy
University of Missouri
Columbia, MO 65211-7010, USA
E-mail: singhdk@missouri.edu
Dr. J. A. Rodriguez-Rivera, Dr. L. W. Harriger
NIST Center for Neutron Research
Gaithersburg, MD 20878, USA
Dr. J. A. Rodriguez-Rivera
Department of Materials Science and Engineering
University of Maryland
College Park, MD 20742, USA

S. Thomas, Prof. A. Ernst
Max-Planck-Institut für Mikrostrukturphysik
Weinberg 2, 06120 Halle, Germany
Dr. T. W. Heitmann
University of Missouri Research Reactor
Columbia, MO 65211, USA
Prof. V. Dugaev
Department of Physics and Medical Engineering
Rzeszów University of Technology
35-959 Rzeszów, Poland
Prof. A. Ernst
Institut für Theoretische Physik
Johannes Kepler Universität
4040 Linz, Austria

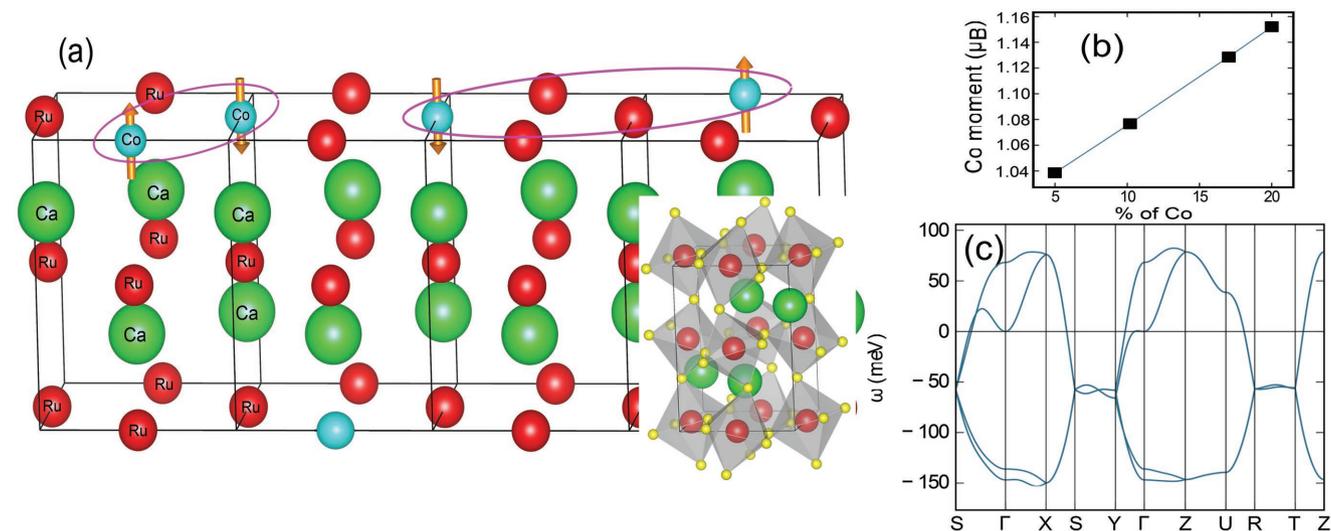

**Figure 1.** Controlled disorder and making of spin-1/2 moment. a) Schematic of random Ru-substitution by Co-ion in CaRuO₃. At the substitution coefficient if $x > 8.3\%$, at least one Ru-ion is replaced by Co-ion in the lattice unit cell, on the average, that are randomly distributed; necessary to create dimers with varying exchange constants. Oval depicts a dimer formation. b) Calculation of the spin density of opposite polarity confirms that the net moment of Co-ion is comparable to the moment of spin-1/2 system. c) Calculated spin wave dispersion relation in Co-substituted CaRuO₃, at $x = 0.2$, confirms the formation of singlet Co dimers, as schematically shown in (a).

calculations show that the substitution of Ru by Co in CaRuO₃ gives rise to quasi-discrete levels due to the hybridization of the localized $t_{2g}$ multiplets of Co atom to the propagating states of O $2p$ electrons in the valence band. In the localized configuration, the Co-ion occupies the low spin state and manifests an effective spin-1/2 configuration. It is confirmed by theoretical estimation of Co-ion moment for different substitution percentages, see Figure 1b. At the chemical substitution of 8.3% to 17%, at least one Ru site per unit cell is replaced by the Co-ion. At higher substitution, while the underlying orthorhombic structure is still intact (see Figure S1, Supporting Information), low spin state of Co-ion occupies two or more basis sites per unit cell, on the average, that are randomly distributed. This random distribution of effective spin-1/2 generate multiple partitioning effects in dimer formation that involve nearest as well as the further separated spins, as shown in the schematic description in Figure 1a. The calculated spin dispersion relation for a Co-substituted compound, Ca(Co₀.₂Ru₀.₈)O₃, exhibits four modes due to four Co atoms in the unit cell (see Figure 1c). The exchange interaction between magnetic moments is calculated using the magnetic force theorem. It is implemented within the KKR Green's function method via following equation[13]

$$J_{ij} = \frac{1}{4\pi} \int_{\infty}^{\varepsilon_F} d\varepsilon \operatorname{Im} \operatorname{Tr}_L \left[ \Delta_i(\varepsilon) G_{\uparrow}^{ij}(\varepsilon) \Delta_j(\varepsilon) G_{\downarrow}^{ji}(\varepsilon) \right] \quad (1)$$

where $\operatorname{Tr}_L$ denotes the trace over the angular momentum, $G^{ij}(\varepsilon)$ is the back scattering operator of spin $\sigma$ between the sites $i$ and $j$ and $\Delta_i(\varepsilon)$ ($= t_{\uparrow}^i - t_{\downarrow}^i$) is defined by the single scattering matrices that are closely related to the exchange splitting corresponding to the magnetic atom $i$. Two of the calculated modes in Figure 1c are almost degenerate, indicating the formation of the dimers. The interaction between the nearest neighbor Co moments, also the dominant interaction, was found to be antiferromagnetic in nature with the estimated exchange strength of $J = -5.98$ meV. The dispersion relation demonstrates the absence of any magnetic order at low temperatures. Thus, such dimer–dimer interactions manifest magnetic properties that are usually observed in a two-dimensional frustrated lattice of spin-1/2.

## 2. Results and Discussion

### 2.1. Experimental Characterization of Ca(Co$_x$Ru$_{1-x}$)O₃

The controlled chemical substitution of Ru by Co ion in the non-Fermi liquid paramagnetic metal CaRuO₃ induces metal-to-insulator transition as a function of the substitution coefficient $x$, as shown in **Figure 2**a. The metal–insulator transition progresses gradually until $x = 0.15$. A dramatic change in the electrical properties is observed across $x = 0.15$ substitution. The electrical resistance is found to obey the Arrhenius-type activation in compounds with $x > 0.15$. Thus, it confirms the insulating characteristic of the system where the underlying physics is appropriately described by the Heisenberg formulation.[14,15] Theoretical calculations, as shown in Section 2.2, also show a significant decrease in the density of states as Ru-ions are replaced by Co-ion. Magnetic susceptibility measurements on a characteristic substitution coefficient, $x = 0.2$, confirms the paramagnetic character to the lowest measurement temperature (Figure 2b).[16] In Figure 2c, we show the Curie–Weiss fit to the high temperature susceptibility of few substitution coefficients in the insulating regime. The Curie–Weiss temperature, $\Theta_{CW}$, increases from −144.9 to −220 K as $x$ varies from 0 to 0.2. The enhancement in $\Theta_{CW}$ is much more pronounced in compounds with $x > 0.15$, i.e., when the Co ions occupy more than one site per unit lattice. The average antiferromagnetic interaction $J_{Co-Co} = -6.1$ meV, estimated from $\Theta_{CW}$ using the mean field approximation for Co–Co interaction only in composition $x = 0.19$, is consistent with the fitting of the susceptibility data using a more detailed mean field analysis involving both Co–Co and Ru–Ru interaction, see Figure 2d (see the Supporting Information for mean field formulation). For the fitting purposes, the Curie–Weiss constant for Ru sublattice was fixed to the stoichiometric value of 0.55, as reported in ref. [11]. The estimated exchange constants, $J_{Co-Co} \simeq -5.6$ meV and $J_{Ru-Ru} \simeq 0.19$ meV, point to the dominant role of Co–Co interaction in $x > 0.15$ compounds. In stoichiometric compound CaRuO$_3$ ($x = 0$), $J_{Ru-Ru}$ is larger than the value estimated in the presence of Co-ions.[11] Introduction of Co seems to alter the exchange interaction between Ru–Ru ions. We also tried to estimate $J_{Co-Co}$ for the stoichiometric fixed value of $J_{Ru-Ru}$. However, fixing $J_{Ru-Ru}$ in the mean field approximation did not yield good fit to the susceptibility data. Reasonably accurate estimations of $J_{Co-Co}$ and $J_{Ru-Ru}$, in the modified environment where Ru ions are substituted by Co ions, required the floating of both exchange constants in the fitting of susceptibility data.

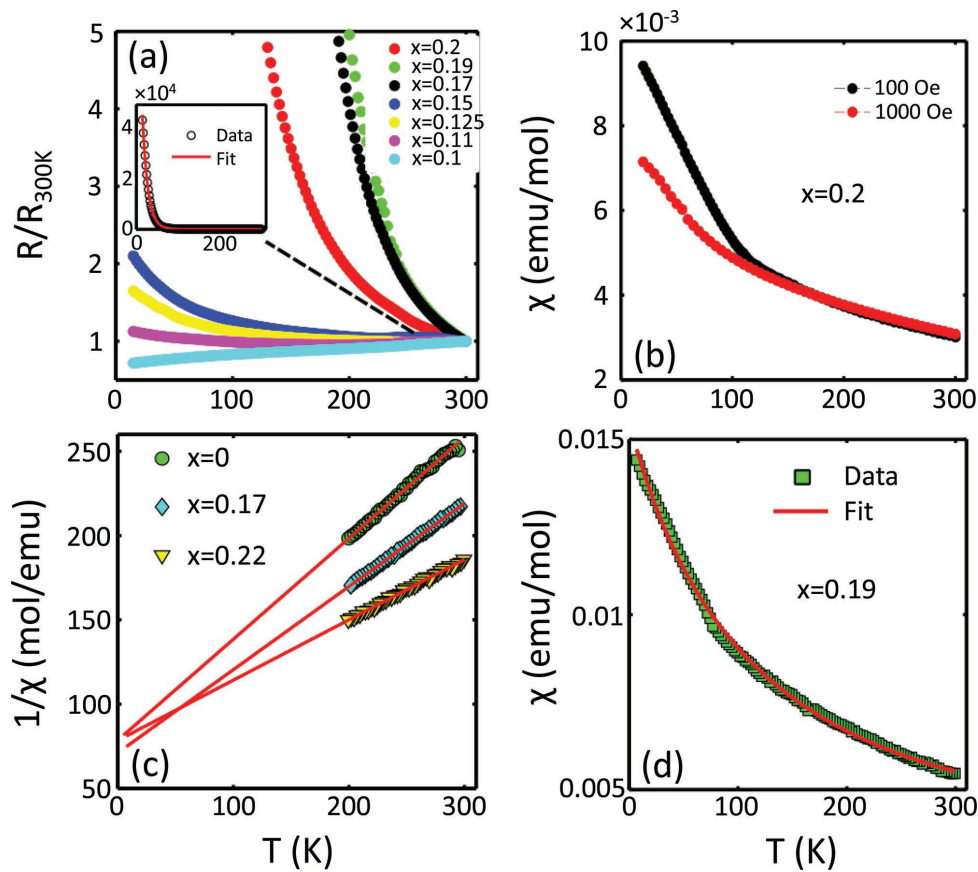

**Figure 2.** Metal-insulator transition and the Curie-Weiss fit to magnetic susceptibility. a) Plot of electrical resistivity versus temperature for various $x$. As $x$ increases, a transition from metal to insulator phase is clearly observed. At $x > 0.15$, the system enters the activated insulating regime. Inset shows the Arrhenius formulation-type activation in $x = 0.19$. b) Magnetic susceptibility measurements confirm the weak paramagnetic character of Co-doped insulator Ca(Co$_{0.2}$Ru$_{0.8}$)O$_3$. Similar behavior is observed in other composition with $x > 0.15$. c) The Curie-Weiss fit to the susceptibility data reveal a significant enhancement in $\Theta_{CW}$ as $x$ crosses over in the strong insulator regime. d) Mean-field fit to magnetic susceptibility data, using $J_{Co-Co}$ and $J_{Ru-Ru}$ exchanges, in a characteristic composition $x = 0.19$.

Despite a reasonably large exchange constant, no static magnetic order was detected in elastic neutron scattering measurements on any of the Co-substituted CaRuO$_3$ (see Figure S4, Supporting Information). Elastic measurements were also performed on cold spectrometer SPINS with high resolution setup to a low temperature of $T = 0.3$ K in both zero and applied field of $H = 7$ T. Yet, no signature of the development of any type of magnetic order is observed. Further investigation of the absence of magnetic order is performed using heat capacity measurements in both zero and applied field. Once again, no anomalous signature in the heat capacity data, indicating the absence of any type of magnetic order in zero or applied field (see Figure S3, Supporting Information). Investigation of possible dynamic correlation is performed using detailed ac susceptibility and inelastic neutron scattering measurements. The ac susceptibility measurements, with a frequency range of 500–10$^4$ Hz, reveal strong frequency-dependent dynamic susceptibility in Co substituted CaRuO$_3$. The characteristic plots of simultaneously measured static ($\chi'$) and dynamic ($\chi''$) susceptibilities as a function of temperature are shown in **Figure 3** (see Figure S2 for other $x$, Supporting Information). A small cusp is observed to develop at $T \simeq 70$ K in $\chi'$ as $x$ increases. However, the overall increase in $\chi'$ at low temperature is marginal compared to the value at $T = 300$ K. Thus, the system exhibits weak paramagnetic character at low temperature; also consistent with Figure 2b. Unlike $\chi'$, $\chi''$ exhibits dramatic reversal in its temperature dependence across the critical substitution $x = 0.15$—from being strong at high frequency at low temperature for $x < 0.15$ to being strong at high temperature for $x \geq 0.15$. For instance, in $x = 0.17$, $\chi''$ is much stronger at room temperature compared to the almost negligible magnitude at low temperature. The completely surprising observation in $\chi''$ at high $x$ can be ascribed to two possibilities: first, the fluctuation at high temperature is $Q$-independent, which coalesce to finite wavevectors in the momentum–energy ($Q$–$E$) space as temperature is reduced. At high temperature, those localized excitations melt in the broader realm of the $Q$–$E$ space. Such a scenario will give rise to the continuum spectrum in $Q$–$E$ space at high temperature. Alternatively, the system fluctuates at much faster rate at low temperature and, therefore, cannot be detected by the ac susceptibility probe.

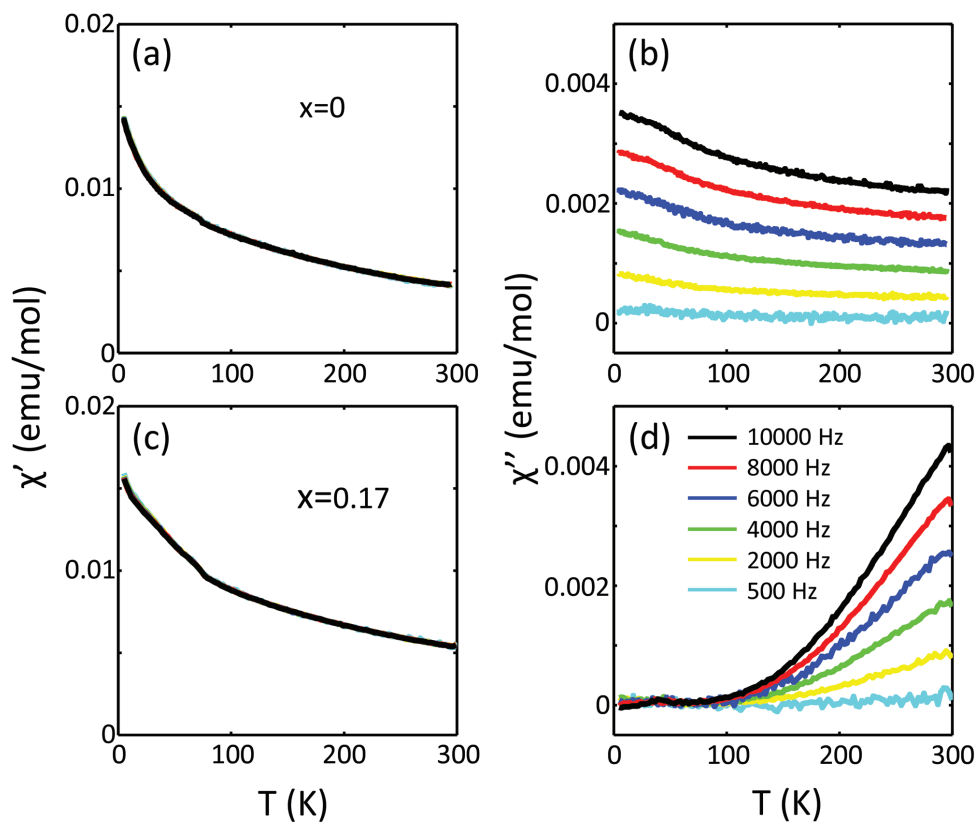

**Figure 3.** Temperature dependences of $\chi'$ (static) and $\chi''$ (dynamic susceptibility). a,c) Temperature dependence of $\chi'$ at different frequencies (500–$10^4$ Hz) at $H_{ac}$ = 10 Oe for characteristic substitution coefficients, $x$, in Ca(Co$_x$Ru$_{1-x}$)O$_3$ (also see Figure S2, Supporting Information). Weak paramagnetic behavior is detected in all $x$. b,d) $\chi''$ at different ac frequencies are plotted as a function of temperature for characteristic $x$ values (see Figure S2 for other substitution coefficients). Unlike $\chi'$, $\chi''$ exhibits strong frequency dependence. $\chi''$ becomes more prominent at higher temperature in the insulating regime.

## 2.2. Theoretical Interpretation of Magnetic Properties of Ca(Co$_x$Ru$_{1-x}$)O$_3$: Resonant Levels of Co Impurities

To elucidate the experimental finding, we performed first-principle calculations using full-potential Green-function method within the density functional theory.[17–19] Co impurities in the CaRuO$_3$ host were efficiently modeled within a coherent potential approximation (CPA), which is specially designed to describe disorder effects of the long-range order.[20,21] In addition, we have applied an embedded cluster method to take account of short-range effects.[22] In this approach, a cluster with impurity atoms and the environment is incorporated into the host matrix in the real space representation. Our tests showed that both approaches, the CPA and the embedded cluster method envisaged for Co in CaRuO$_3$, yield similar outcomes.

The results of ab initio calculations show that the doping of CaRuO$_3$ with Co demonstrates the formation of quasi-discrete levels in the valence band (as shown in **Figure 4**). They are due to hybridization of the localized states of Co atom (the states of $t_{2g}$ multiplet) to the propagating states of electrons in the valence band. The energy of the quasi-local state falls into continuous energy spectrum of the valence band, which can be the reason of a strong resonant scattering of electrons when the Fermi level is close to the energy of quasi-discrete resonant state (see Figure 4). At a certain Co concentration, this leads to a localization of the electrons at the scattering centers, in this case Co ions. In Co dimers, these states can form hydrogen like molecules with the spin of 1/2.

The strong scattering of electrons from the resonant impurities makes it possible to reach the criterion of Anderson localization due to disorder. One can use a simplified model to calculate the electron scattering time and estimate a possibility of the Anderson localization.

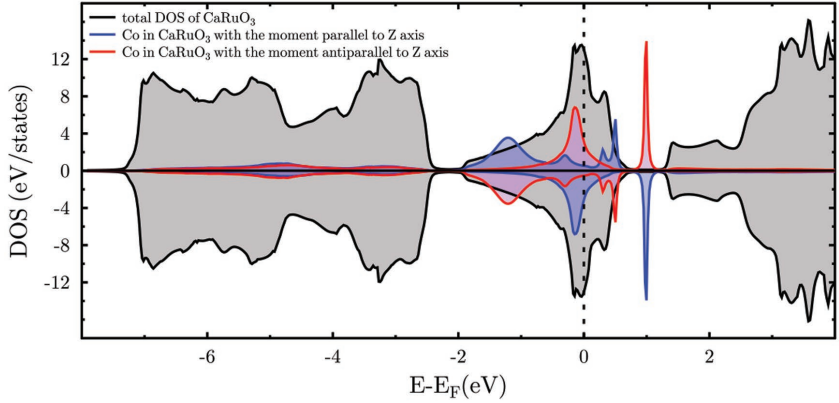

**Figure 4.** Electron density of states in CaRuO$_3$ with additional peaks of quasi-discrete levels created by Co atoms with magnetic moments parallel and antiparallel to the $Z$ axis in the continuous spectrum of host material.

The Hamiltonian of electrons in the valence band hybridized to impurity level $\varepsilon_i$ reads

$$H = \sum_{\mathbf{k}} \varepsilon_{\mathbf{k}} a_{\mathbf{k}}^{\dagger} a_{\mathbf{k}} + \frac{V}{\sqrt{N_0}} \sum_{\mathbf{k}i} \left( a_{\mathbf{k}}^{\dagger} a_i + h.c. \right) + \sum_i \varepsilon_i a_i^{\dagger} a_i \quad (2)$$

where $\varepsilon_{\mathbf{k}}$ is the electron energy spectrum in the valence band (we take $\varepsilon_{\mathbf{k}} \simeq \hbar^2 k^2 / 2m$ near the band edge), $\varepsilon_i$ is the energy of resonant level, $N_0$ is the total number of host atoms (Ru) in the lattice, $V$ is the matrix element of hybridization, which we assume to be constant due to the screening of Co potential. The creation and annihilation operators $a_{\mathbf{k}}^{\dagger}$, $a_{\mathbf{k}}$ correspond to electrons in the valence band. Correspondingly, $a_i^{\dagger}$ and $a_i$ are for electrons localized at Co atom at $\mathbf{r} = R_i$.

The self-energy of electrons due to the resonant scattering

$$\Sigma_e = \frac{xV^2}{\varepsilon - \varepsilon_i + i\Gamma_i} \quad (3)$$

where $x = N_i/N_0$, $N_i$ is the number of impurities (Co atoms) in the sample and $\Gamma_i(\varepsilon)$ determines the width of a single resonant level.

The imaginary part of $\Sigma_e$ determines the electron density of states and the relaxation time $\tau_e$ of electrons with energy $\varepsilon$

$$\frac{\hbar}{2\tau_e} \equiv \Gamma_e = \frac{xV^2\Gamma_i}{(\varepsilon - \varepsilon_i)^2 + \Gamma_i^2} \quad (4)$$

In its turn, the level width $\Gamma_i$ can be also calculated from the self-energy of the localized electron

$$\Gamma_i = -\frac{V^2\Omega}{N_0} \operatorname{Im} \int \frac{d^3\mathbf{k}}{(2\pi)^3} \frac{1}{\varepsilon - \varepsilon_{\mathbf{k}} + i\Gamma_e} \quad (5)$$

$$= \frac{V^2\Gamma_e}{n_0} \int \frac{v_0(\varepsilon_{\mathbf{k}})d\varepsilon_{\mathbf{k}}}{(\varepsilon - \varepsilon_{\mathbf{k}})^2 + \Gamma_e^2} \quad (6)$$

where $n_0 = N_0/\Omega$ is the volume density of Ru atoms in the host, $\Omega$ is the volume of crystal, $v_0(\varepsilon_{\mathbf{k}})$ is the electron density of states in the valence band without impurities (one can take $v_0(\varepsilon_{\mathbf{k}}) \simeq m^{3/2}\varepsilon_{\mathbf{k}}^{1/2}/\sqrt{2}\pi^2\hbar^3$ near the band edge and calculate integral (6) from 0 to $\infty$).

Equations (4) and (6) determine both $\Gamma_i$ and $\Gamma_e$ as a function of energy $\varepsilon$. The electron localization takes place when the Fermi energy $\varepsilon_F \leq \Gamma_e(\varepsilon_F)$. Since $\Gamma_e$ is proportional to concentration of Co, the criterion $\varepsilon_F = \Gamma_e(\varepsilon_F)$ allows to estimate the critical value of $x$.

The dependence $\Gamma_e(\varepsilon)$ was calculated numerically using Equations (4) and (6) for $m = m_0$ (the mass of free electron) and various values of $V$. This dependence shows a resonance near $\varepsilon = \varepsilon_i$. Since the Fermi level is located within the resonance, the scattering of Fermi electrons from Co impurities is strong.

The dependence of $\Gamma_e(\varepsilon_F)$ on Co concentration $x$ is presented in **Figure 5**. The electron states with $\Gamma_e \geq \varepsilon_F$ are localized. This shows that at $x \simeq 0.12$–$0.15$ occurs the metal-insulator transition due to the Anderson localization of electron states at the Fermi level.

## 3. Inelastic Neutron Scattering Measurements

Detailed inelastic neutron scattering (INS) measurements were performed at cold spectrometers MACS and SPINS, using final neutron energies of $E_f = 5$ meV and 5 meV at which the spectrometers resolution were $\simeq 0.32$ and 0.28 meV, respectively. In **Figure 6**a–f, we plot the background corrected and thermally balanced $Q$–$E$ spectra at few characteristic temperatures in zero magnetic field (also see Figure S5, Supporting Information). Two types of trends are immediately observed in the Q-E maps: first, at low temperature of $T = 80$ mK, a well-defined gapped excitation at $E = 5.9$ meV develops at $Q = 1$ Å$^{-1}$. The dynamic structure factor is much broader than the instrument resolution in both momentum and energy space (Figure 6a–c). Also observable is a much weaker excitation at $Q = 2$ Å$^{-1}$ at the same energy, which follows Co form factor. Given the fact that no magnetic order was detected in any of these materials, the observation of localized excitation is completely surprising. The excitation is non-dispersive in nature. No such localized excitation was observed in CaRuO$_3$ or Ca(Co$_x$Ru$_{1-x}$)O$_3$ with low substitution coefficient (see Figure S7, Supplementary Materials).[11] It further suggests that the gapped excitation arises due to the Co$^{4+}$-Co$^{4+}$ interaction. As temperature increases, the excitation gradually weakens before becoming indistinguishable from the remnant background.

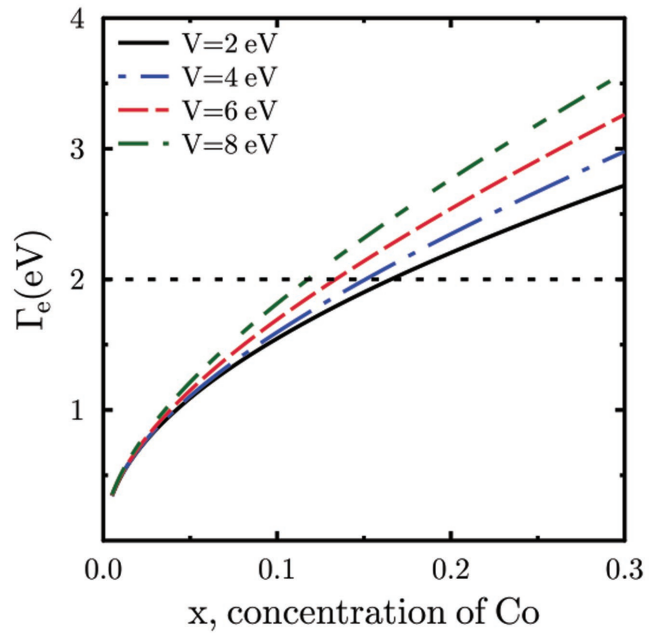

**Figure 5.** Variation of the electron relaxation rate $\Gamma_e(\varepsilon_F)$ with concentration of Co atoms for different values of hybridization parameter $V$. We take the energy $\varepsilon_F = 2$ eV. This calculation shows that the localization criterion is satisfied within a relatively small range of $x$ for different values of $V$.

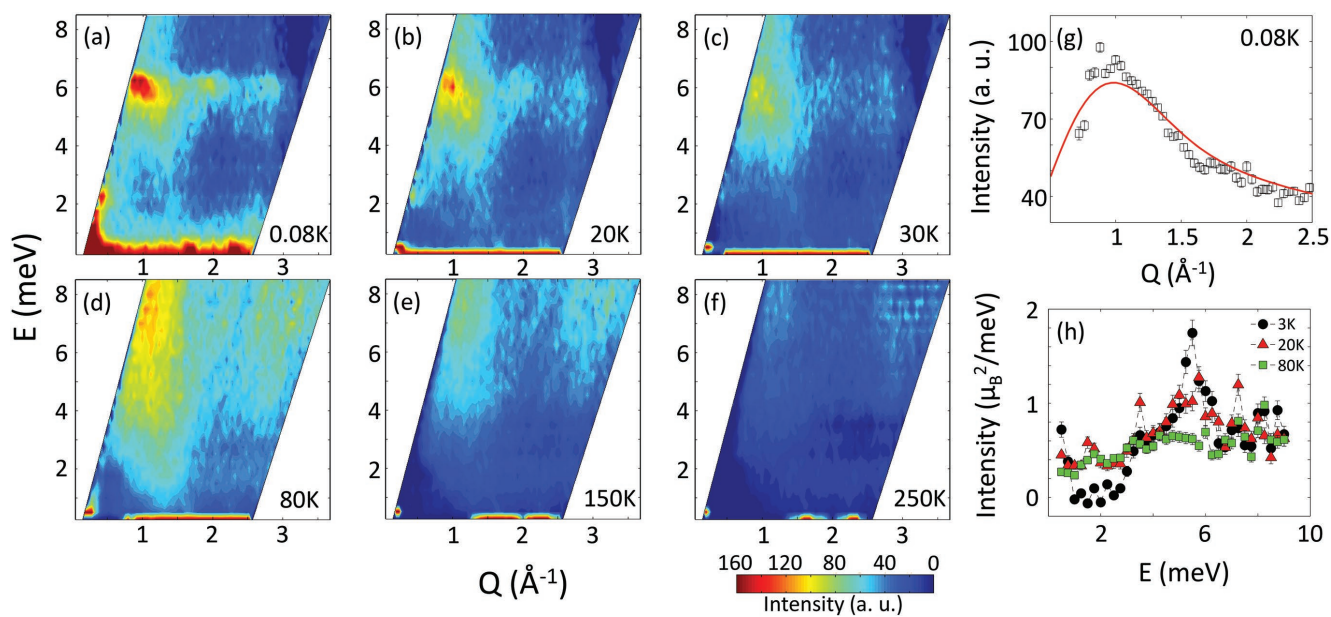

**Figure 6.** Localized excitation from dimer–dimer interaction to quasi-continuum spectrum. a–c) Energy–momentum ($Q$–$E$) map, depicting localized excitation in Ca(Co$_{0.17}$Ru$_{0.83}$)O$_3$, at different temperatures, ranging from $T = 80$ mK to 30 K, are plotted. Experimental data are background corrected and thermally balanced by multiplying with the Bose factor, $(1-\exp(-E/k_BT))$. A well-defined excitation is found to be centered at $E = 5.9$ meV and $Q = 1$ and 2 Å$^{-1}$, albeit weakly. It follows the form factor of Co. The excitation becomes weaker as temperature increases. d–f) As temperature increases further, the excitation reappears. However, it is much more broader in both momentum and energy, reminiscent of a quasi-continuum behavior. At $T = 80$ K, the quasi-continuum fills the majority of the $Q$–$E$ space, which persists to an unusually high temperature of $T = 250$ K. g) At $T = 80$ mK, the averaged intensity versus $Q$ is well described by a singlet-to-triplet transition due to the dimer-dimer interaction between nearest and further away spins. (h) Measurements on SPINS at $Q = 1.15$ Å$^{-1}$ provide more quantitative perspective.

The collective gapped excitation at low temperature (Figure 6a–c), centered at $E = 5.9$ meV, is comparable to the mean field exchange constant of Co–Co interaction, $J_{Co-Co} \simeq 6$ meV. The broad width of the inelastic peak suggests a continuum of excitations, involving exchange constants beyond nearest neighbors. To understand the origin of this collective excitation, we plot the averaged intensity of the inelastic spectra between $E = 4.5$ and 7.5 meV as a function of the wave-vector $Q$ at $T = 80$ mK. As shown in Figure 6g, the experimental results are well described by the singlet-to-triplet excitations of spin dimers, involving nearest and the next nearest neighbor interactions[23,24]: $I(Q) = |F(Q)|^2 \sum m_i^2 [1 - \sin(Qd_i)/(Qd_i)]$, where $m_i^2$ is the squared moment per formula unit and $d_i$ is distance between nearest or next nearest neighbor. Simple inclusion of the nearest neighbor interaction does not fit the data well.[24,25] The estimated values of $m_i^2$ and $d_i$ at $T = 80$ mK are: $m_1^2 = 0.367$ $\mu_B^2$, $d_1 = 3.85$ Å, $m_2^2 = 0.15$ $\mu_B^2$ and $d_2 = 5.4$ Å. The near neighbor distances, $d_1$ and $d_2$, are remarkably close to the lattice values. The estimated dimer moments are also consistent with similar observations in the chemically ordered spin-1/2 systems, such as the Kagome structure.[10] As temperature increases, the dynamic structure factor gradually decreases and disappears above $T > 30$ K.

Second, the excitation spectra re-emerges near the same wave-vectors but centered at nominally higher energy $E \simeq 7.8$ meV, as temperature is further increased to $T = 80$ K (Figure 6d–f). Most importantly, the high temperature excitations are gapless and occupy a significant portion of the $Q$–$E$ space. The broad continuous tail in both energy and the momentum space is reminiscent of the quasi-continuum spectrum, observed in geometrically frustrated lattice of spin-1/2.[9,26] The quasi-continuum excitation at high temperature complements the ac dynamic susceptibility measurements in x > 0.15 (Figure 3, and Figure S2, Supporting Information). As we cross into the Heisenberg regime, i.e., x > 0.15, the dynamic susceptibility pattern changes significantly. The fluctuation is now confined to higher temperature only (see Figure 3 and Figure S2, Supporting Information). It confirms the gapped nature of the excitation at low temperature and the continuum spectrum at higher temperature. Although, the spectral weight is distributed in the large region of $Q$–$E$ space, the remnants of localized excitations at $Q = 1$ and 2 Å$^{-1}$ can still be seen, albeit much weaker, at higher temperature. It further suggests that the quasi-continuum is arising due to the breaking of the dimer pairs. These excitations persist to very high temperature, $T \simeq 250$ K, with reduced scattering intensity. This process becomes complicated due to the presence of spin-orbital coupling in the host matrix of $CaRuO_3$.[27] Perhaps, this can be one of the reasons that the quasi-continuum persists to such a high temperature.[28] Because of their distinct temperature dependencies, these phenomena are considered magnetic.[26,29]

The experimental results, manifested by the MACS data in Figure 6a–f, are confirmed by independent measurement on cold spectrometer SPINS. A set of two-dimensional inelastic scans at $Q = 1.15$ Å$^{-1}$ at different temperatures are plotted in Figure 6h. The $Q$-value for the SPINS measurements were intentionally selected to avoid the inelastic peak position and thus, obtain a more realistic estimation of the spectral weight distribution to the background as functions of temperature and energy. Measurement at low temperature, $T = 3$ K, confirms the gapped nature of the localized excitation. As temperature increases, the localized excitation gradually melts and the spectral weight is transferred to low energy. It is as if the dimer pairs break and the freely fluctuating spins occupy the low energy spectrum between the singlet ground state and the triplet excited state in a continuous fashion, thus making it a gapless continuum. At higher temperature, $T \geq 80$ K, the spectral weight also relocates to higher energies. This observation is consistent with the neutron data in Figure 6d–f, where significant scattering is observed at $T \simeq 80$ K. Above $T > 80$ K (also comparable to the peak value of the broad fluctuation spectrum, 7.8 meV), thermal fluctuation starts dominating the spin fluctuations. At $T = 250$ K, it becomes very weak. Measurements in applied magnetic field of $H = 10$ T did not reveal any field dependence of the inelastic properties of the system (see Figure S6, Supporting Information).

## 4. Conclusion

Our experimental results provide new perspective to the study of quantum magnetism in spin-1/2 system for three reasons: first, the absence of any type of magnetic order rules out the ordered dimer arrangement of artificial spin-1/2, ultimately responsible for the VBS state. Similarly, the disorder in the lattice would prohibit coherent spin fluctuation, hence, the VBL state. However, the dimer formation in the insulating composition of $Ca(Co_xRu_{1-x})O_3$ involves simultaneous superposition of nearest neighbor and further separated artificial spin-1/2, pointing in random directions, by virtue of disorder. The dimer–dimer interaction exhibits gapped singlet-to-triplet transition. Second, the break away spins of dimer pairs at high temperature create freely fluctuating fractionalized spins that occupy large cross section of the energy-momentum space. We also note that the fluctuation spectrum persists to much lower energy, as found in the ac susceptibility measurements where the dynamic susceptibility exhibits stronger frequency-dependence at higher temperature. Thus, the localized gapped excitation evolves into a gapless quasi-continuum spectrum as a function of temperature.[30–32] These two phenomena combined with the absence of magnetic order in $Ca(Co_xRu_{1-x})O_3$, where $0.15 < x < 0.2$, qualify to be a new quantum magnetic state, which depicts remarkable similarity to the resonant valence bond state. Additionally, this quasi-continuum spectrum persists to very high temperature,

$T \simeq 250$ K. The persistence of quasi-continuum to such an unusually high temperature extends the investigation to the semiclassical regime, which is a new frontier in the study of the quantum-mechanical properties in magnetic systems. Third and most important, we successfully demonstrate that a combination of disorder and the non-frustrated lattice can provide a new platform for future researches on quantum magnetism via the creation of local artificial spin-1/2.

## 5. Experimental Section

First-principles calculations of electronic and magnetic properties of $Ca(Co_xRu_{1-x})O_3$ were performed using a self-consistent Green function method within the density functional theory. The high purity polycrystalline samples of $Ca(Co_xRu_{1-x})O_3$ were synthesized by conventional solid state reaction method using ultrapure ingredients of CoO, $RuO_2$, and $CaCO_3$. Starting materials were mixed in stoichiometric composition, pelletized, and sintered at 950° for 3 d in oxygen-rich environment. The furnace-cooled samples were grinded, pelletized, and sintered at 1000° for another 3 d. Samples were intentionally synthesized at slightly lower temperature and for longer duration to preserve the oxygen stoichiometry. Resulting samples were characterized using Siemens D500 powder X-ray diffractometer (XRD), confirming the single phase of material, see Figure S1 in the Supporting Information. The X-ray diffraction data were analyzed using the FullProf suite for the Rietveld refinement, confirming the high quality single phase of materials (see Figure S1, Supporting Information). As shown in Figure S1 (Supporting Information), every single peak of the XRD pattern is identified with the orthorhombic structure of $Ca(Co_xRu_{1-x})O_3$. Four-probe technique was employed to measure electrical properties of $Ca(Co_xRu_{1-x})O_3$ using a closed-cycle refrigerator cooled 9 T magnet with measurement temperature range of 1.5–300 K. Detailed ac susceptibility measurements were performed using a Quantum Design Physical Properties Measurement System with a temperature range of 2–300 K.[33] Elastic neutron scattering measurements were performed on the pristine powder samples of $Ca(Co_xRu_{1-x})O_3$ at the thermal triple-axis spectrometer TRIAX at MURR and triple-axis spectrometer SPINS at the NIST Center for Neutron Research with fixed final energies of 14.7 and 5 meV, respectively. Inelastic measurements were performed on cold spectrometers MACS and SPINS with fixed final neutron energies of 5 and 5 meV at which the spectrometers' energy resolutions were determined to be $\simeq 0.32$ and 0.28 meV, respectively.

## Supporting Information

Supporting Information is available from the Wiley Online Library or from the author.

## Acknowledgements

The research at MU is supported by the Department of Energy, Office of Science, Office of Basic Energy Sciences under the grant no. DE-SC0014461. This work utilized neutron scattering facilities supported by Department of Commerce. S.T. and A.E. acknowledge the "Collaborative Research Center Sonderforschungsbereich SFB 762, Functionality of Oxide Interfaces." The work of V.K.D. was supported by the National Science Center in Poland, research project No. DEC-2012/06/M/ST3/00042. D.K.S. envisaged the research idea and supervised every aspect of the experimental research. J.G., A.D., and Y.C. contributed equally to this work. Samples were synthesized by J.G. and A.D. The measurements were carried out by J.G., A.D., Y.C., L.H., J.R., T.H., and D.K.S. Data analysis was carried out by J.G., Y.C., and D.K.S. Theoretical calculations were carried out by S.T., V.G., and A.E. The paper was written by D.K.S. and A.E. where everyone contributed.

## Conflict of Interest

The authors declare no conflict of interest.

## Keywords

Anderson localization, perovskites, singlet-triplet transitions, quantum magnetism